\documentclass[conference,a4paper,twocolumn]{IEEEtran}

\IEEEoverridecommandlockouts

\ifCLASSINFOpdf
\else
\fi
  \usepackage{tipa}
  \usepackage{graphicx}
	\usepackage{epsfig}
	\usepackage{epstopdf}
	\usepackage{upgreek}
	\usepackage[nolist,nohyperlinks]{acronym}

\usepackage[cmex10]{amsmath}

\usepackage{color, soul}

\usepackage{subfigure}

\usepackage{cite}

\begin{acronym}
\acro{NoC}{Network-on-Chip}
\acro{WNoC}{Wireless Network-on-Chip}
\acro{EM}{Electromagnetic}
\acro{SiO$_2$}{Silicon Dioxide}
\acro{AIN}{Aluminum nitride}
\acro{SiP}{System-in-Package}
\acro{SDM}{Software-defined metamaterial}
\end{acronym}

\begin{document}



\title{Opportunistic Beamforming in\\ Wireless Network-on-Chip}


%
\author{
\IEEEauthorblockN{
Sergi Abadal\IEEEauthorrefmark{1}, Adri\'{a}n Marruedo\IEEEauthorrefmark{1}, Antonio Franques\IEEEauthorrefmark{3}, Hamidreza Taghvaee\IEEEauthorrefmark{1},\\
Albert Cabellos-Aparicio\IEEEauthorrefmark{1}, Jin Zhou\IEEEauthorrefmark{2}, Josep Torrellas\IEEEauthorrefmark{3}, Eduard Alarc\'{o}n\IEEEauthorrefmark{1}
 }
\IEEEauthorblockA{
\IEEEauthorrefmark{1}\small{NaNoNetworking Center in Catalunya (N3Cat), Universitat Polit\`{e}cnica de Catalunya (UPC), Barcelona, Spain%
}}
\IEEEauthorblockA{
\IEEEauthorrefmark{2}\small{Department of Electrical and Computer Engineering, University of Illinois at Urbana-Champaign (UIUC), Illinois, USA%
}}
\IEEEauthorblockA{
\IEEEauthorrefmark{3}\small{Department of Computer Science, University of Illinois at Urbana-Champaign (UIUC), Illinois, USA}\\%
Email: abadal@ac.upc.edu
}
}




\maketitle


\begin{abstract}
Wireless Network-on-Chip (WNoC) has emerged as a promising alternative to conventional interconnect fabrics at the chip scale. Since WNoCs may imply the close integration of antennas, one of the salient challenges in this scenario is the management of coupling and interferences. This paper, instead of combating coupling, aims to take advantage of close integration to create arrays within a WNoC. The proposed solution is opportunistic as it attempts to exploit the existing infrastructure to build a simple reconfigurable beamforming scheme. Full-wave simulations show that, despite the effects of lossy silicon and nearby antennas, within-package arrays achieve moderate gains and beamwidths below 90\textsuperscript{o}, a figure which is already relevant in the multiprocessor context.
\end{abstract}



%
\IEEEpeerreviewmaketitle




\acresetall

\section{Introduction} \label{sec:introduction} 
Network-on-Chip (NoC) has become the \emph{de facto} standard for the interconnection of cores in multicore processors. However, as we enter the manycore era, the communication requirements increase up to a point where conventional NoCs alone may not suffice \cite{Jerger2017}. Their limited scalability is in fact turning communication into the performance bottleneck of manycore systems, thus calling for new solutions at the interconnect level \cite{Bertozzi2014}.

\begin{figure}[!t]
\centering
\includegraphics[width=0.9\columnwidth]{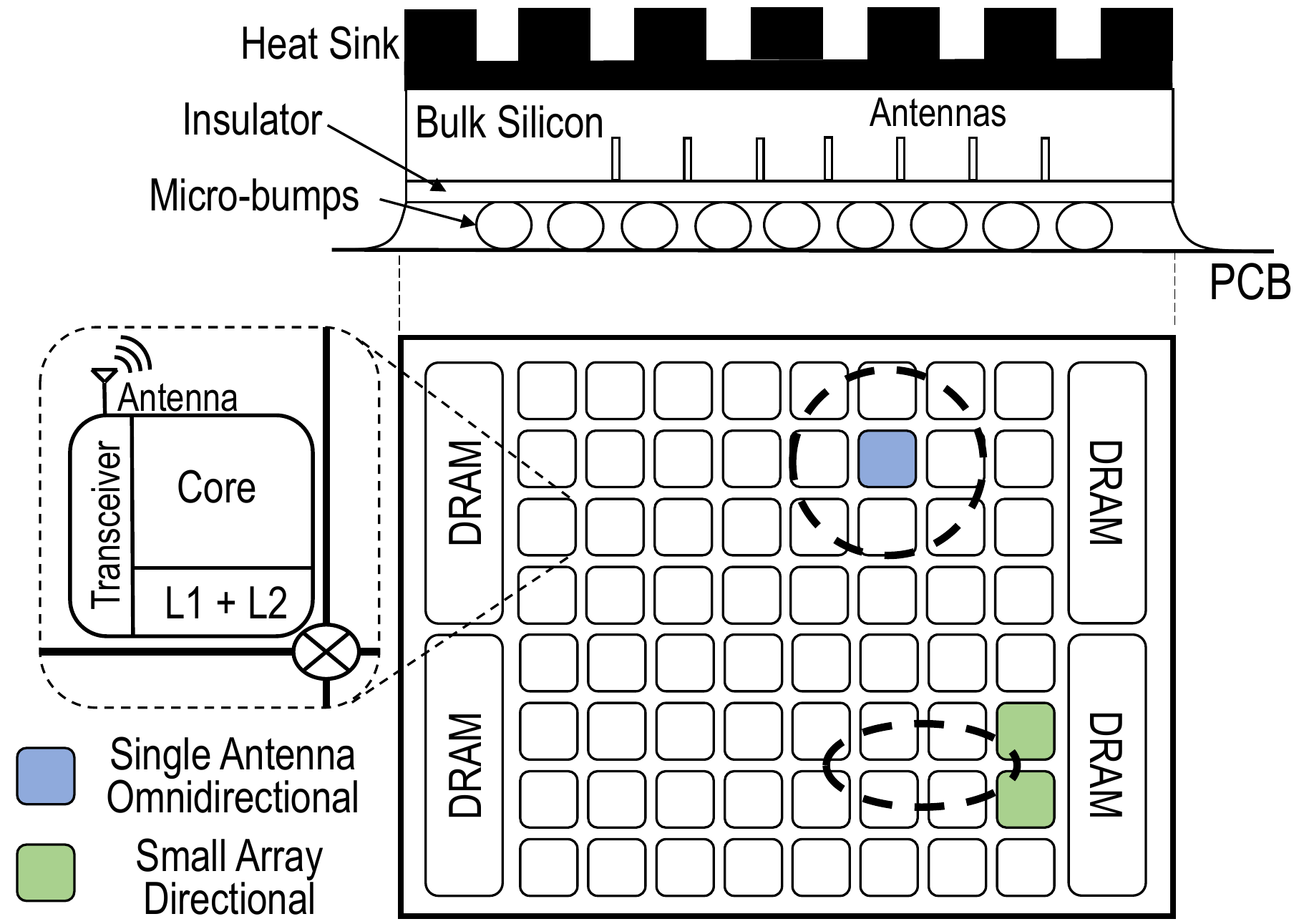}
\vspace{-0.1cm}
\caption{Cross-section and planar view of a multicore processor with a wireless on-chip network. Thanks to the proposed architecture, antennas can operate in isolation (blue, omnidirectional) or form small arrays (green, directional).}
\vspace{-0.2cm}
\label{fig:summary}
\end{figure}

Advances in integrated antennas \cite{Markish2015, Cheema2013} and transceivers \cite{Laha2015, Ahmed2018} have led to the proposal of Wireless Network-on-Chip (WNoC) as a complement of or alternative to existing NoCs \cite{Matolak2012}. As shown in Figure \ref{fig:summary}, WNoCs basically consist of the co-integration of RF front-ends with cores or clusters of cores. Information can be thus modulated and radiated, and radiated signals then propagate through the computing package until reaching the intended destinations. The main advantage of this approach is that distant cores can communicate with low latency thanks to the speed-of-light propagation. In fact, communication is naturally broadcast as long as antennas are roughly omnidirectional. Further, the lack of additional wires between cores provides system-level flexibility not achievable with other technologies. 

From the network architecture perspective, one can distinguish between a large set of WNoC proposals that deploy multiple point-to-point wireless links over a wired NoC \cite{DiTomaso2015, Mansoor2015a, Rezaei2016} and, to a lesser extent, broadcast-based WNoCs \cite{Deb2013, Abadal2018}. On-chip antennas used in these proposals are generally variants of printed dipoles \cite{Zhang2007, Narde2018} or vertical monopoles using through-silicon vias (TSVs) \cite{Wu2017b, Rayess2017, Timoneda2018b}, with rather omnidirectional radiation patterns. As a result, MAC protocols or multiplexing methods are required to avoid collisions and interference in the WNoC \cite{Abadal2018a, DiTomaso2011, Vijayakumaran2014}. However, this approach has important limitations because the number of non-overlapping frequency, code, or time slotted channels achievable in this resource-constrained scenario is relatively small. 

An alternative or complement to the multiplexing schemes mentioned above would be \emph{spatial multiplexing} as proposed in some works \cite{Zhao2008, Mondal2016, Mineo2016, Pano2017, Gade2018}. By using directional antennas, several wireless point-to-point links can coexist in the same frequency-time window and increase the overall available throughput. The main downturn of this approach, however, is that the antennas need to be carefully aligned and that the established links cannot be reconfigured, thereby losing the system-level flexibility and inherently broadcast appeal of the WNoC paradigm. This issue could be partially overcome by means of dynamic beamforming, but this would require the use of antenna arrays in each wireless interface as proposed in \cite{Liu2016}, which is clearly unaffordable given the evident area limitations of the manycore scenario.
 
\begin{figure*}[!ht]
\centering
\vspace{-0.2cm}
\subfigure[Single antenna \label{fig:arrays1}]{\includegraphics[width=0.25\textwidth]{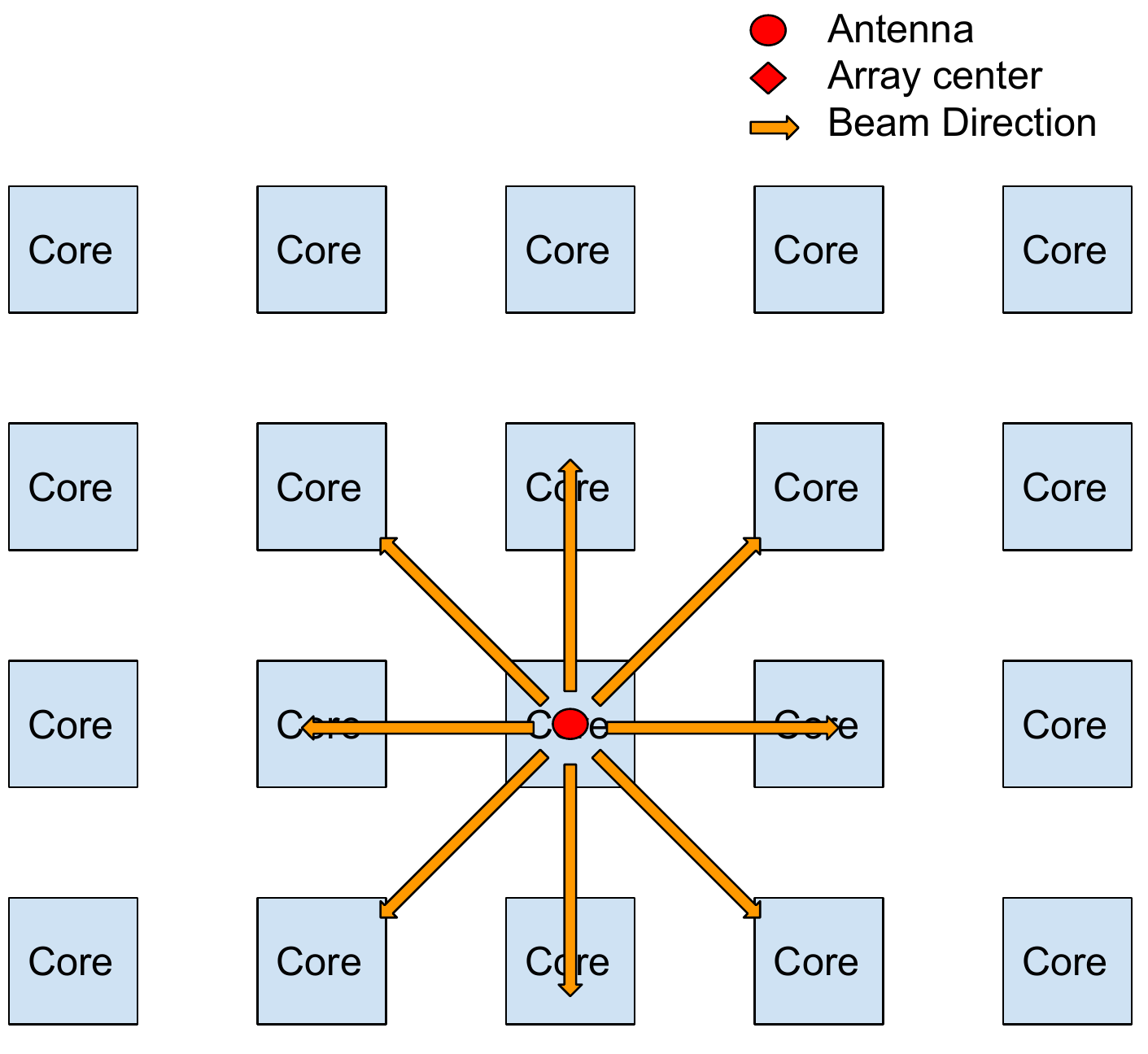}}
\hspace{0.5cm}
\subfigure[Shared array (multi-chip) \cite{Baniya2018a}\label{fig:arrays2}]{\includegraphics[width=0.25\textwidth]{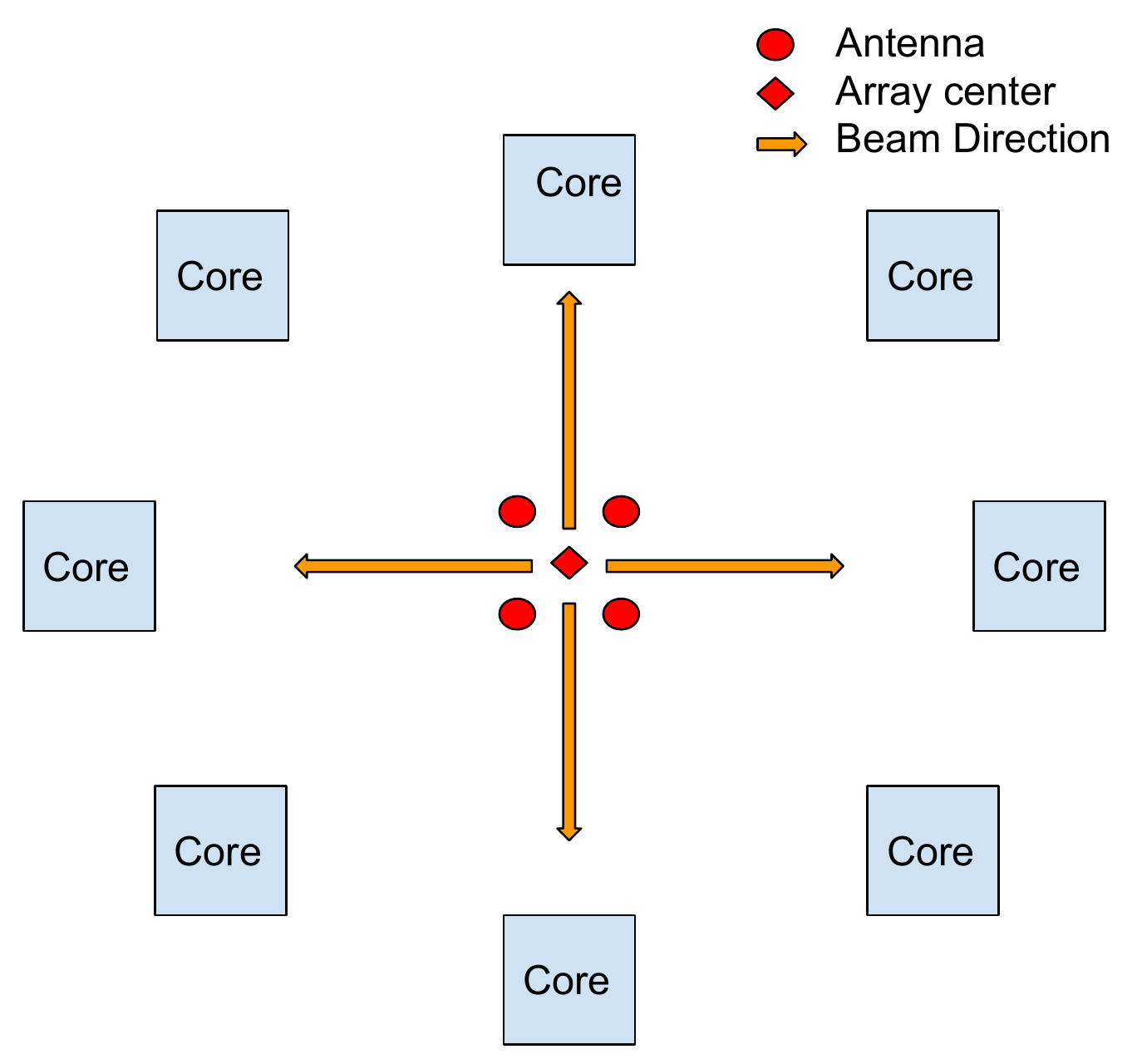}}
\hspace{0.5cm}
\subfigure[Proposed solution\label{fig:arrays3}]{\includegraphics[width=0.25\textwidth]{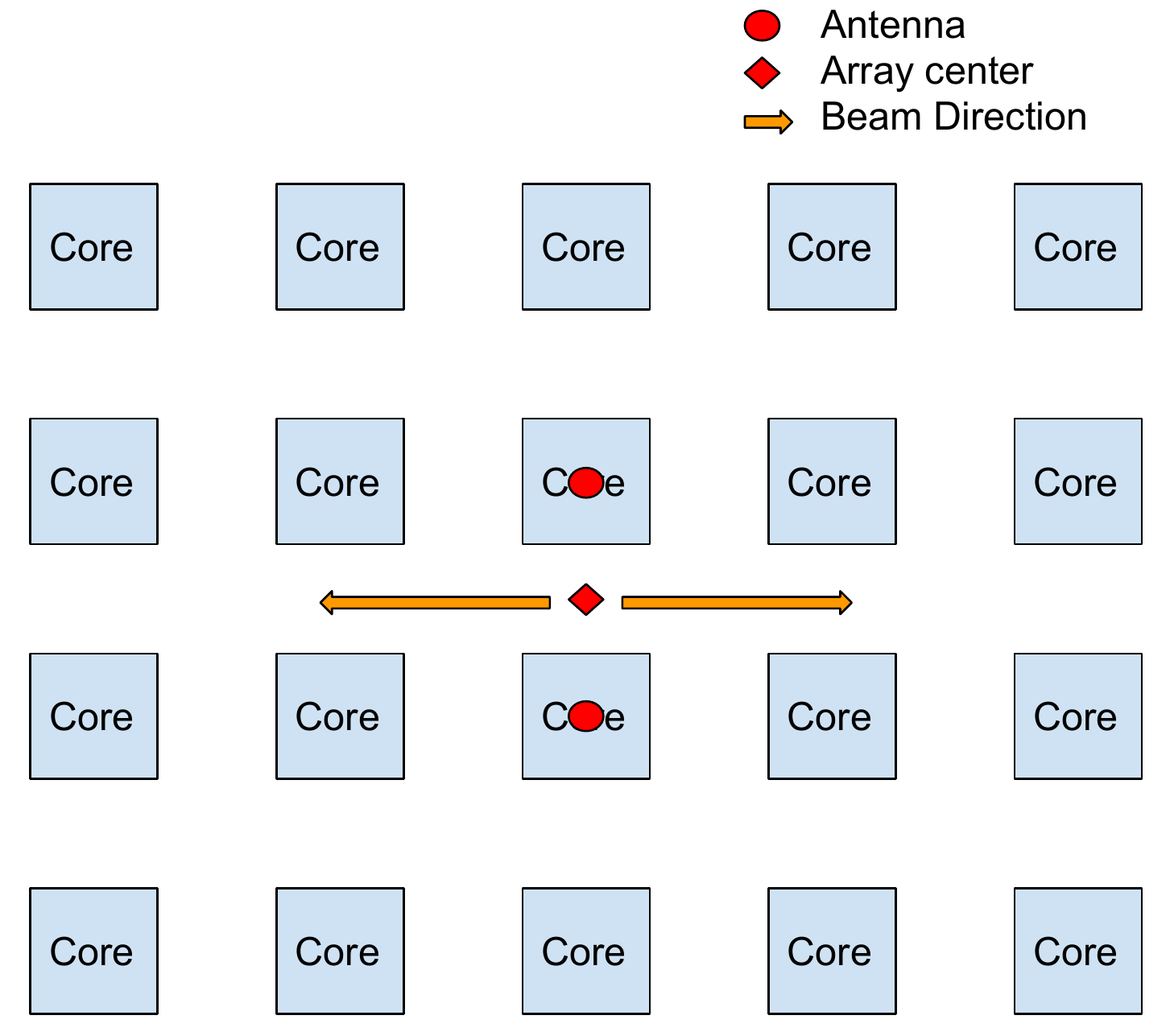}}
\vspace{-0.2cm}
\caption{Summary of antenna configurations in a wireless on-chip network.}
\vspace{-0.4cm}
\label{fig:arrays}
\end{figure*}

This paper proposes an opportunistic solution to this problem, seeking to implement spatial multiplexing in a flexible and affordable way for wireless on-chip networks. The main idea is to leverage the known channel characteristics \cite{Timoneda2018ADAPT}, the already existing high density of on-chip antennas, and the already existing tight synchronization among cores to create small antenna arrays as shown in Fig. \ref{fig:summary}. The proposed solution incurs into small overhead as it only adds a very simple phase shifter to each antenna and an array controller per each group of $n$ antennas. With our scheme, the system can create directional arrays and modify their structure on demand driven by the communication needs of the particular application being run, or simply remain omnidirectional.


The remainder of this paper is organized as follows. Section \ref{sec:desc} presents an overview of the idea and details a potential implementation. Section \ref{sec:arrays} analyzes the theoretically formable patterns, which are later evaluated via full-wave simulations in Section \ref{sec:eval}. Finally, Section \ref{sec:conclusions} concludes the paper.



\section{Opportunistic Beamforming within a Chip Package}
\label{sec:desc}
The great majority of WNoC works consider the collocation of antennas and transceivers either to (groups of) cores \cite{Abadal2018} or to selected routers \cite{Deb2013, DiTomaso2015, Mansoor2015a, Rezaei2016}. In these cases, schematically represented in Fig. \ref{fig:arrays1}, each antenna operates in isolation with a rather broad beam and must be carefully integrated to avoid undesired coupling effects. This, however, restricts the number of wireless interfaces and limits the potential of WNoC in manycore processors. 

Very few works have explored the possibility of actually leveraging coupling to create arrays in chip-scale environments. Only Baniya \emph{et al.} have proposed the integration of small arrays for beam switching in chip-to-chip communication. Their scheme, shown in Fig. \ref{fig:arrays2}, considers that groups of cores share a four-element array that can switch between different radiation directions depending on the location of the receiver. The arrays are built by design rather than opportunistically taking advantage of existing antennas, which complicates the layout and reduces the overall flexibility. Moreover, their work assumes an unconventional chip package. 

Next, we provide an overview of our proposal in Section \ref{sec:overview}, to then discuss the architecture in Section \ref{sec:arch}.

\subsection{Overview of the idea}
\label{sec:overview}
We propose to take advantage of the already existing high density of antennas and, with minor changes, provide means for beamforming within a chip package. The scheme assumes that each core (or group of cores) has its own antenna. By default, each antenna operates in isolation and can be tuned to radiate omnidirectionally to create a \emph{broadcast channel} as depicted in Fig. \ref{fig:arrays1}. When needed, two or more antennas are activated simultaneously and form a small array that delivers a \emph{multicast channel} through directional radiation as shown in Fig. \ref{fig:arrays3}. A controller synchronizes the transceivers to ensure that the constructive interference among antennas results into the desired directional radiation.

The solution is opportunistic and may be even regarded as partially distributed as:
\begin{itemize}
\item It exploits already existing antennas.
\item Cores are, by definition, tightly synchronized by means of a global clock common to the whole processor.
\item Data may be already present in several cores either due to existing architectural mechanisms \cite{AbadalASPLOS, Fernando2019} or enforced by software.
\item It admits a few (architecturally relevant) radiation directions, easy to derive given the destination address.
\end{itemize}
In support of this last argument, it is worth noting that parallel programming libraries include collective primitives that are used in a variety of fundamental algorithms and that generate all-to-all communication patterns \cite{grama2003introduction}. In a conventional mesh NoC, collectives are generally performed within all cores of the same row first, and then within all cores of each column (or vice versa) \cite{Jerger2008, Krishna2011}. Therefore, row/column communication patterns can be architecturally relevant for WNoCs in manycore systems.

 %


\begin{figure}[!ht]
\centering
\vspace{-0.2cm}
\includegraphics[width=0.8\columnwidth]{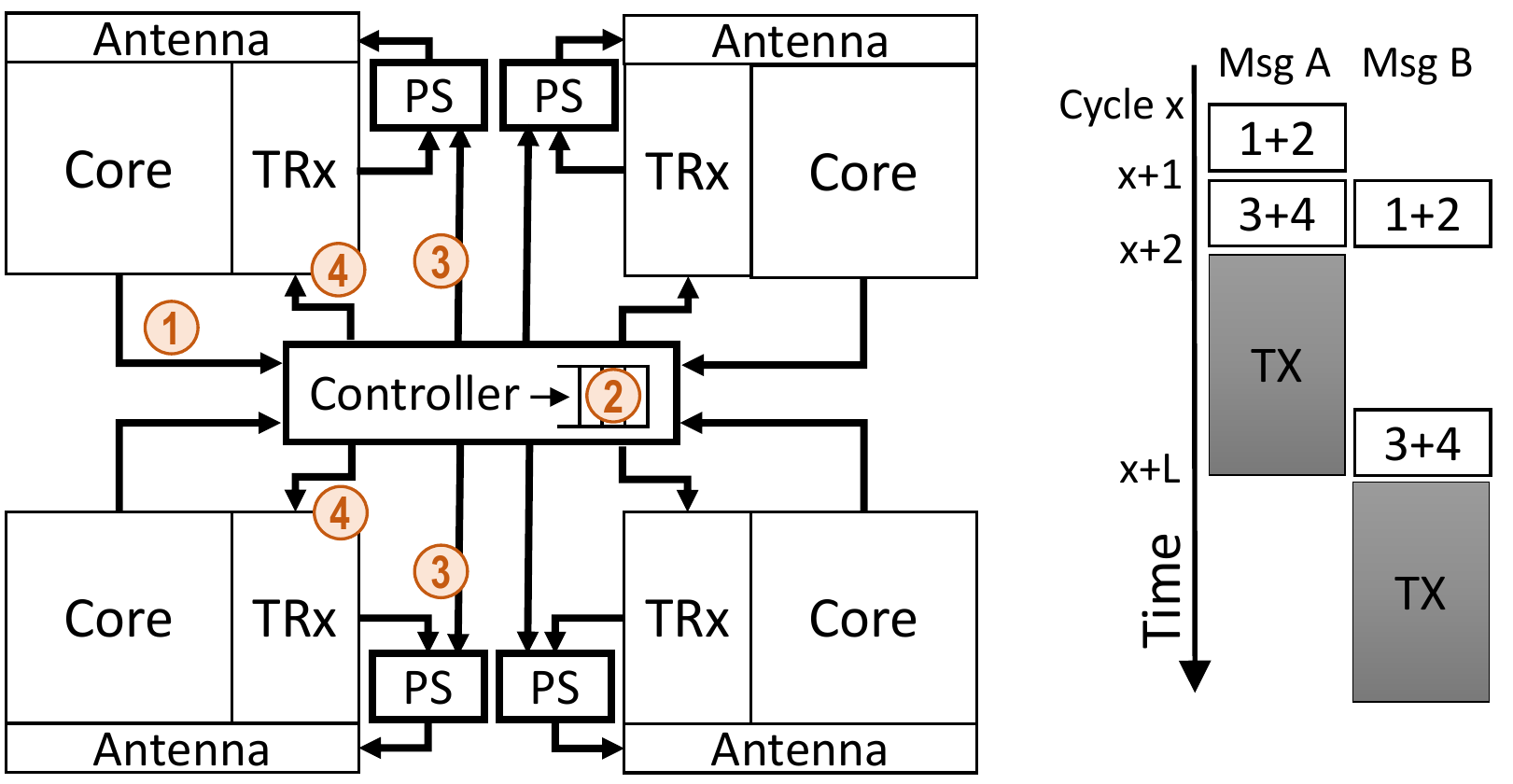}
\vspace{-0.2cm}
\caption{Proposed architecture in a 2$\times$2 cluster of cores (not to scale). Thick lines refer to added components. TRx stands for transceiver, whereas PS stands for phase shifter. Right plot illustrates the timing of the different steps.}
\vspace{-0.4cm}
\label{fig:schematic}
\end{figure}

\begin{figure*}[!ht]
\centering
\vspace{-0.2cm}
\subfigure[0\textsuperscript{o}]{\includegraphics[width=0.16\textwidth]{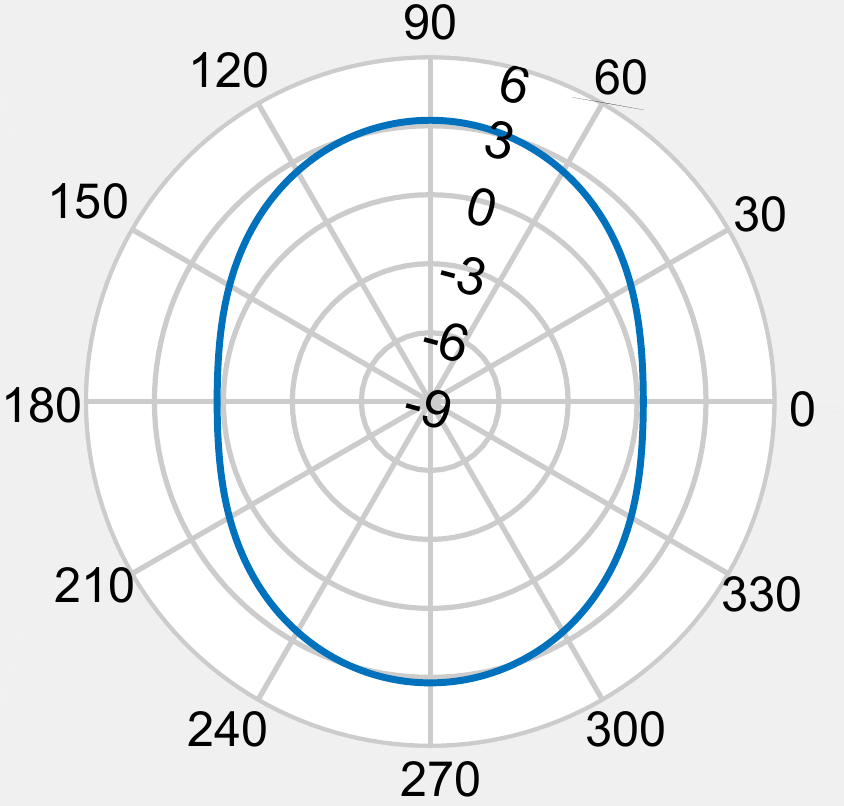}}
\subfigure[90\textsuperscript{o}]{\includegraphics[width=0.16\textwidth]{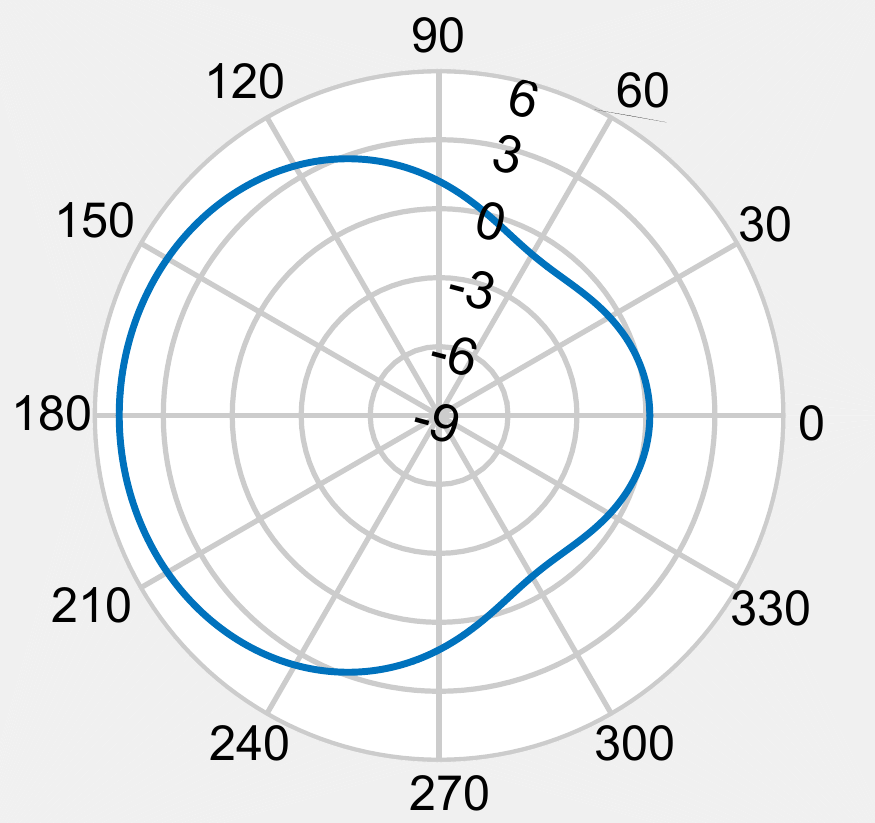}}
\subfigure[180\textsuperscript{o}]{\includegraphics[width=0.16\textwidth]{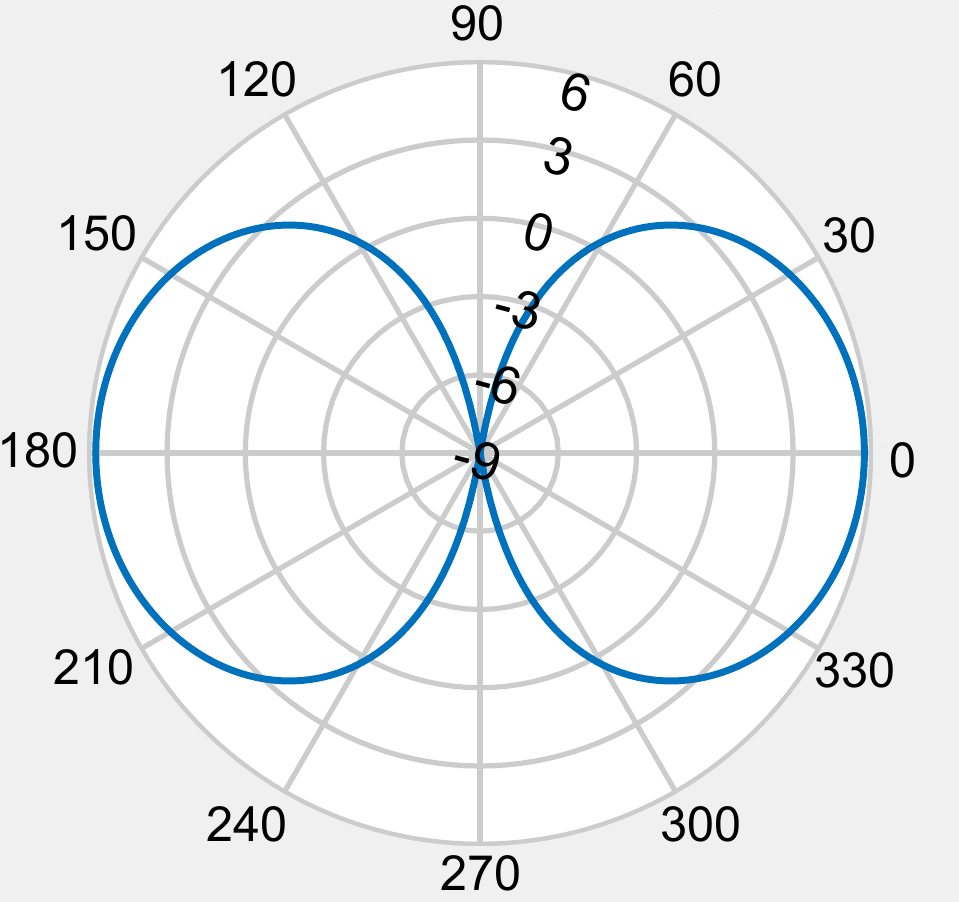}}
\subfigure[270\textsuperscript{o}]{\includegraphics[width=0.16\textwidth]{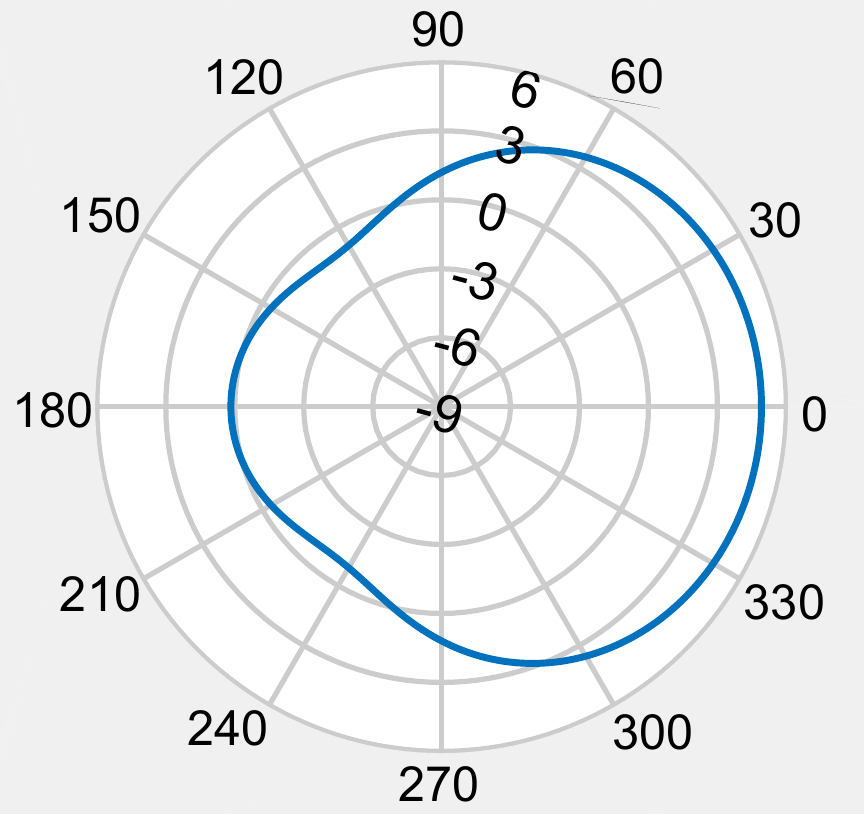}}
\subfigure[Diagonal, 0\textsuperscript{o}]{\includegraphics[width=0.16\textwidth]{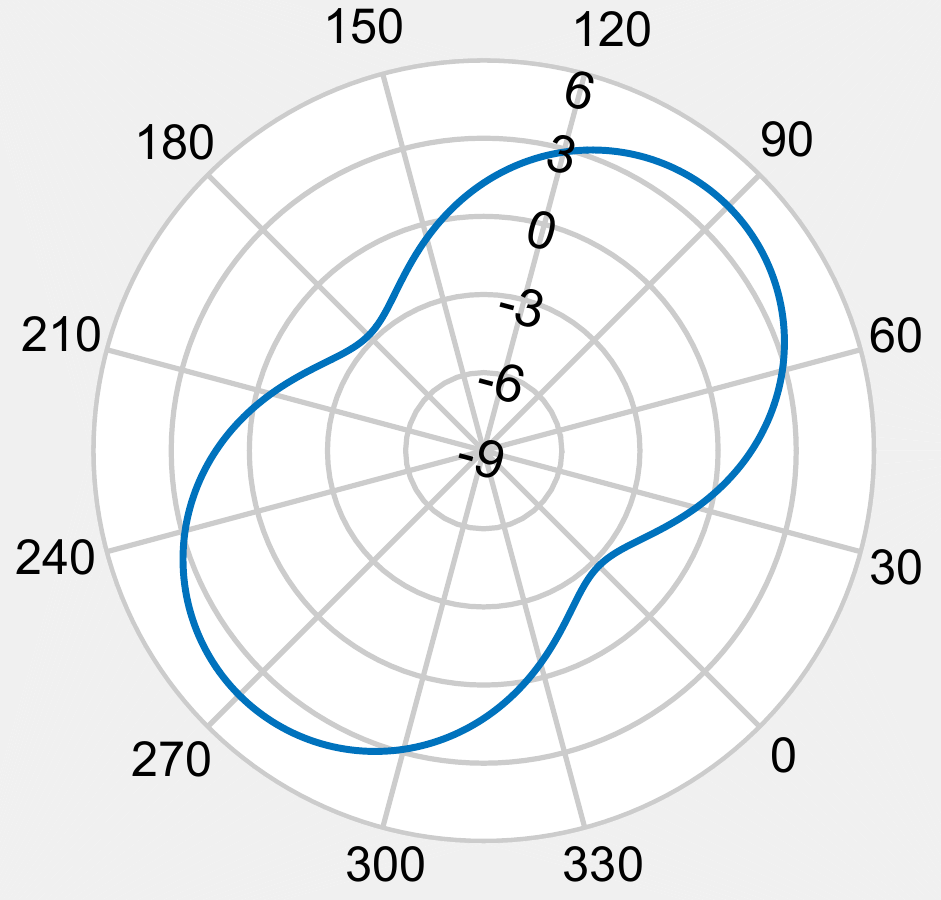}}
\subfigure[Diagonal, 180\textsuperscript{o}]{\includegraphics[width=0.16\textwidth]{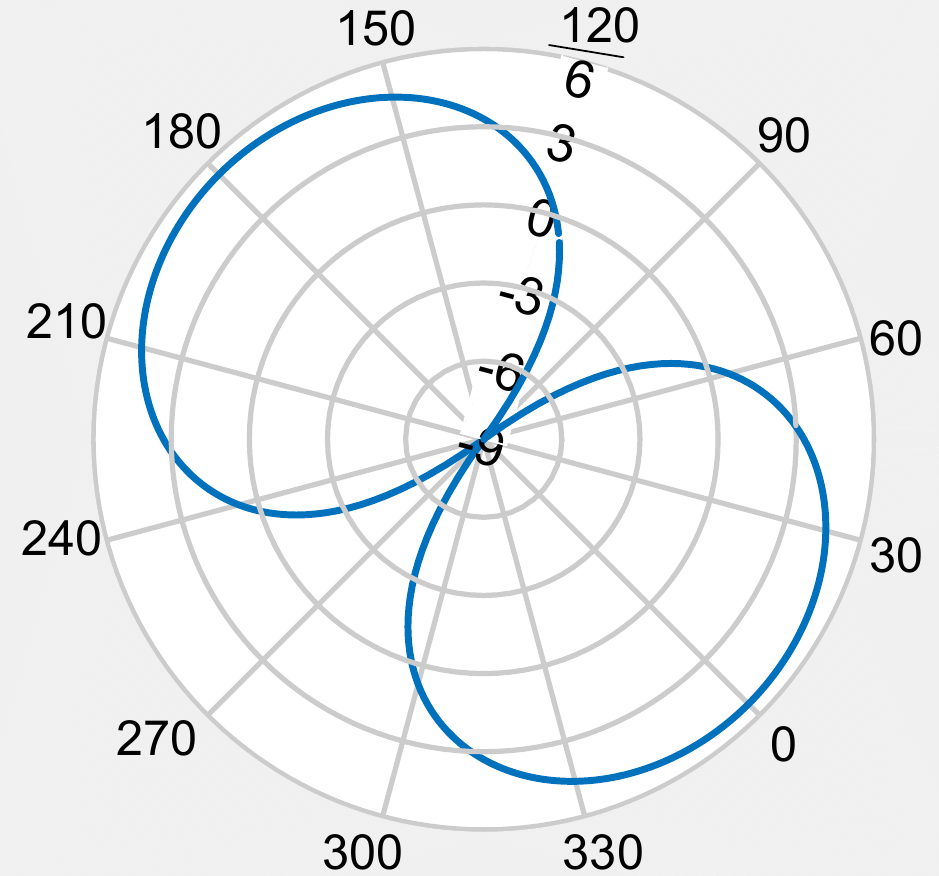}}
\vspace{-0.2cm}
\caption{Theoretical gain patterns in the chip plane for two omnidirectional antennas separated by $d=\lambda/4$ with different phase shifts $\Delta\Phi$.}
\vspace{-0.3cm}
\label{fig:beams}
\end{figure*}

\subsection{Architecture}
\label{sec:arch}
Figure \ref{fig:schematic} shows a schematic representation of our proposed solution, exemplified in a group of 2$\times$2 cores. Each transceiver (TRx) is augmented with a phase shifter (PS) with a very limited number of states (initially set to 0\textsuperscript{o}), whereas a controller is added to each cluster of cores. By default, cores use their omnidirectional antennas in isolation and the controller does not intervene. However, when the creation of a directional channel is required, cores communicate with the controller as depicted in Fig. \ref{fig:schematic}:
\begin{enumerate}
\item The core sends the wireless packet to the controller, which places it into the queue.
\item Upon arrival, the controller checks the destination address and evaluates the best beam direction.
\item Based on the chosen direction, the controller notifies the relevant phase shifters. 
\item While setting the phase shifters, the controller sends the wireless packet to the relevant transceivers, which modulate the information and radiate it.
\end{enumerate}
These steps are analogue to the pipeline stages of a NoC router and, thus, we can use similar timings \cite{Park2012a}. As shown in Fig. \ref{fig:schematic}, steps 1-2 and 3-4 can be performed in the first and second cycle, respectively, in a pipelined fashion. In any case, wireless transmissions are typically longer than two cycles and therefore the controller does not become a bottleneck. 

The beamforming policy enforced by the controller can take different forms. In a first approach, a greedy algorithm could direct the beam as close as possible to the destination, which is feasible since the positions of the controller and the destination are known and static. This, however, could increase the likelihood of collisions if not used judiciously. An alternative is to co-design the controller with the link-layer or network-layer protocols to create non-overlapping spatial channels. In the latter case, beams could be reconfigured every $R$ cycles according to past communication demands \cite{DiTomaso2015}. 


To drive the phase shifters, the controller includes a \emph{beam table} with \{beam direction, phase shift vector\} pairs. The row/column of the controller and the destination are compared, thus determining the beam direction. The number of beam directions is assumed minimal, which implies very simple phase shifters and a small beam table. As we have seen in Section \ref{sec:overview}, such coarse-grained configuration is already architecturally relevant. 

The present architecture can be scaled to larger arrays, and thus sharper beams, to increase the number of spatial channels in chips with many cores and antennas. A hierarchical controller structure or existing architectural/software mechanisms can be exploited to ensure proper data replication and antenna control in dynamic beamforming schemes for WNoC.

\section{Array Formation Analysis}
\label{sec:arrays}
To analyze which are the patterns that can be formed, we resort to fundamental antenna array theory \cite{BalanisBOOK}. We first assume omnidirectional antennas, which in the chip scenario could be achieved with vertical monopoles. It is then considered that antennas are deployed homogeneously with a fixed distance between them, but also that the frequency of operation is a design choice, leaving antenna spacing as a parameter in terms of $\lambda$. We simplify the design space by focusing on short spacings, with the aim of (1) favoring close integration of antennas and (2) avoiding grating lobes appearing when spacing becomes larger than $\lambda$, which could create undesired interferences and complicate the architecture. 

For simplicity, we start with simple two-element arrays and explore several configurations with phase shifts $\Delta\Phi$ of 0\textsuperscript{o}, 90\textsuperscript{o}, 180\textsuperscript{o}, and 270\textsuperscript{o}. The conventional choice is $d=\lambda/2$, which delivers broadside and end-fire patterns with 6 dB and 4.62 dB of peak gain, as well as beamwidths of 60\textsuperscript{o} and 120\textsuperscript{o} for the shifts of $\Delta\Phi=0$ and $\Delta\Phi=180$, respectively (patterns not shown for the sake of brevity). It is therefore a good option for row/column communications, although the flexibility is a bit limited: it does not allow to obtain single-sided patterns.

As a feasible alternative, we considered $d=\lambda/4$ which yields the patterns shown in Figure \ref{fig:beams}. Such scheme offers remarkable single-sided beams for $\Delta\Phi=90$ and $\Delta\Phi=270$, with 4.91 dB of peak gain, 166\textsuperscript{o} beamwidth, and a front-to-back ratio of 4.67 dB. With $\Delta\Phi=180$, the end-fire pattern reduces the beamwidth to 90\textsuperscript{o} and increases the peak gain by 1.11 dB with respect to $d=\lambda/2$ because it matches with the Hansen-Woodyard condition ($\Delta\Phi \approx 2\pi d/\lambda + \pi/n$ with $n=2$) used to optimize end-fire radiation. In the diagonal directions, where $d=\lambda\sqrt{2}/4$, both $\Delta\Phi=0$ and $\Delta\Phi=180$ provide interesting beams with width 92\textsuperscript{o} and 104\textsuperscript{o} and peak gain 4.22 dB and 5.36 dB, respectively. Should the architect need a diagonal one-sided beam, the frequency can be adjusted accordingly to achieve $d=\lambda/4$ in the diagonal direction.

\section{Far-Field Characterization}
\label{sec:eval}
The analysis of Section \ref{sec:arrays} provides interesting, but entirely theoretical design points. We confirm the results through full-wave electromagnetic simulations with CST MWS \cite{CST}. The chip package scheme shown in Fig. \ref{fig:summary} is modeled in CST, including a 11-$\upmu$m layer of silicon dioxide ($\varepsilon=3.9$, lossless) as insulator, a 700-$\upmu$m layer of bulk silicon ($\varepsilon=11.9$ and resistivity $\rho=10\,\Omega\cdot$cm) as substrate, and a 200-$\upmu$m layer of thermal interface material ($\varepsilon=8.6$, lossless). The top and bottom boundaries (heat sink and micro-bumps, respectively) are modeled as perfect electrical conductors, whereas lateral boundaries are considered as perfect matched layers. 

For this study, we choose monopole antennas because most of the power is radiated laterally towards the chip edges. Monopoles can be implemented with TSVs and their length controlled thanks to existing electroplating techniques \cite{Timoneda2018b}. In CST, the monopole is modeled as a thin vertical cylinder through the silicon and the length is optimized to minimize the return loss at 60 GHz. Monopole arrays are placed in the center of a 20$\times$20 chip and are surrounded by more antennas at $\lambda/4$ distance to recreate a high-density WNoC.

\begin{figure}[!t]
\centering
\vspace{-0.2cm}
\subfigure[0\textsuperscript{o}]{\includegraphics[width=0.24\columnwidth]{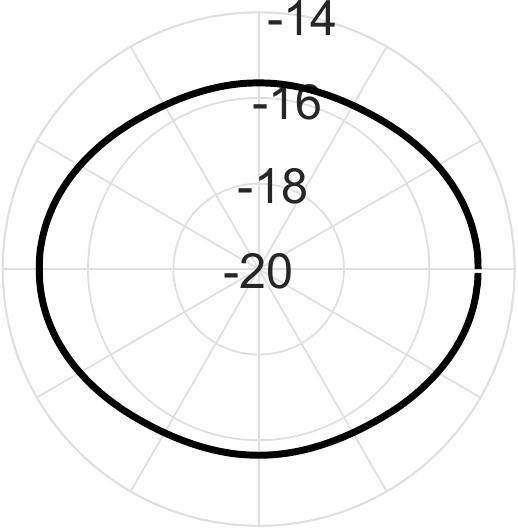}}
\subfigure[90\textsuperscript{o}]{\includegraphics[width=0.24\columnwidth]{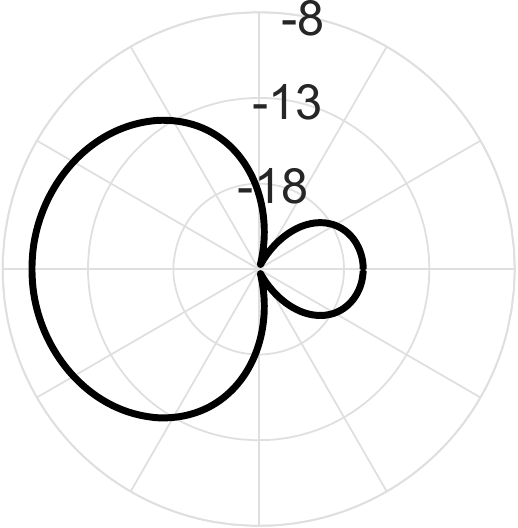}}
\subfigure[180\textsuperscript{o}]{\includegraphics[width=0.24\columnwidth]{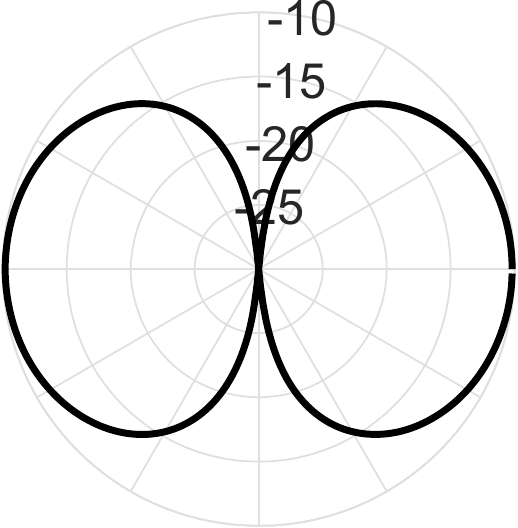}}
\subfigure[Omni]{\includegraphics[width=0.24\columnwidth]{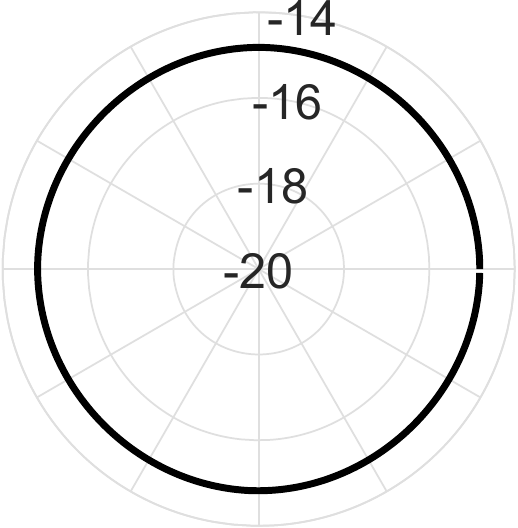}}
\subfigure[0\textsuperscript{o}]{\includegraphics[width=0.24\columnwidth]{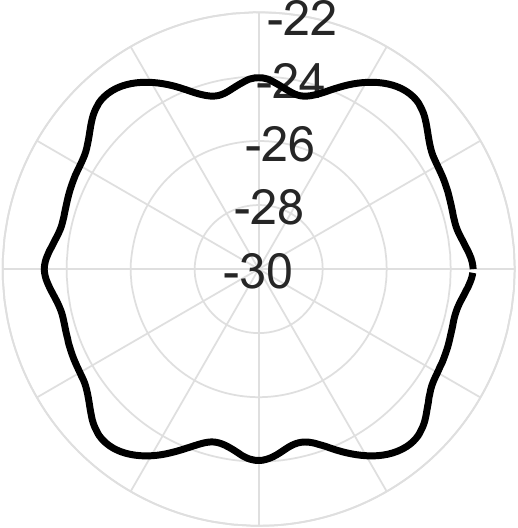}}
\subfigure[90\textsuperscript{o}]{\includegraphics[width=0.24\columnwidth]{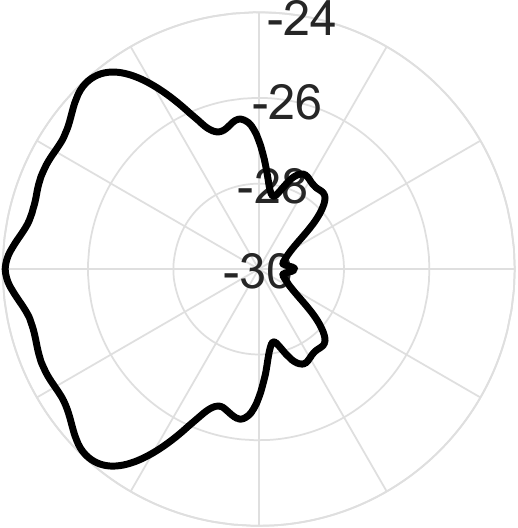}}
\subfigure[180\textsuperscript{o}]{\includegraphics[width=0.24\columnwidth]{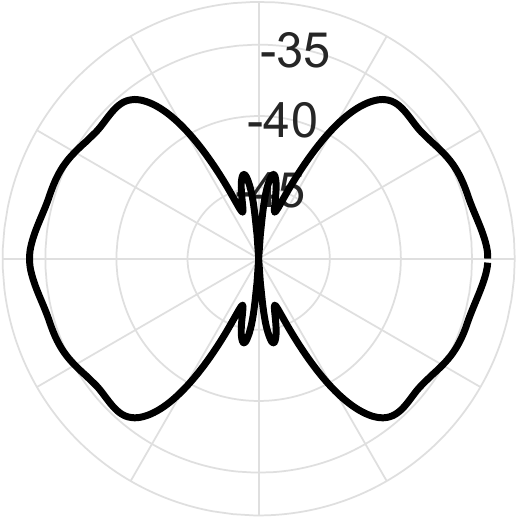}}
\subfigure[Omni]{\includegraphics[width=0.24\columnwidth]{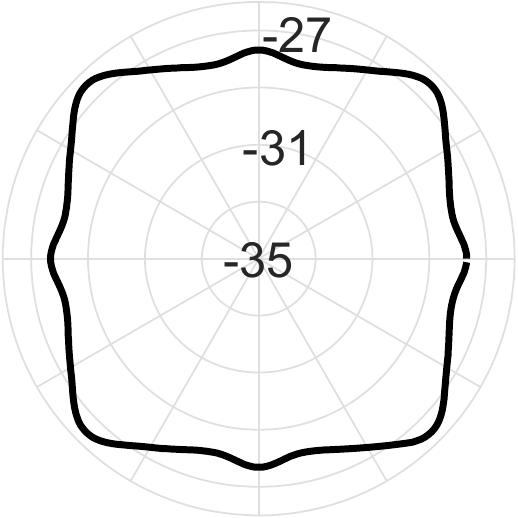}}
\vspace{-0.4cm}
\caption{Radiation patterns of a two-antenna array with separation $\lambda$/4 within a realistic chip package and surrounded by interfering antennas. Patterns (a-c) are evaluated in the near field, whereas patterns (e-g) are evaluated at a distance of 5$\lambda$. Plots (d) and (h) illustrate single-antenna patterns.\label{fig:sim}}
\vspace{-0.4cm}
\end{figure}

\noindent \textbf{Radiation patterns:} we simulate two-antenna array to verify that the patterns analyzed in Sec. \ref{sec:arrays} are possible within the chip environment. We obtain the gain (IEEE) in the azimuthal plane, within the silicon, both near to and far from the array. Results in Fig. \ref{fig:sim}(d) and \ref{fig:sim}(h) show that when a single antenna is excited, the pattern remains roughly omnidirectional even with the presence of interfering antennas around. The main reason is that the coupling between nearby antennas is low given the presence of the lossy silicon between them. This also explains how the theoretical directional patterns can be replicated with reasonable accuracy, as shown in Fig. \ref{fig:sim}(a-c) and \ref{fig:sim}(d-f). Near the array ($\sim$$\lambda$/2), the lossy silicon leads to a reduction of the gain by $\sim$15 dB in average. Far from the array ($\sim$5$\lambda$), we observe that the gain decreases sharply for $\Delta\Phi=180$, to the point of discouraging the use of this radiation mode. We speculate that this is due to the presence of reflections coming from the ground plane or the heat sink.

\noindent \textbf{Scaling trends:} we simulate linear arrays with three and four antennas to evaluate the potential of the proposed approach when scaled. Table \ref{tab:scalability} compares several alternatives with phase shifts $\Delta\Phi$ of 0\textsuperscript{o}, 90\textsuperscript{o}, or following the Hansen-Woodyard condition. The gain is measured at distance of 5$\lambda$. The omnidirectional mode improves in terms of gain, whereas the end-fire mode with $\Delta\Phi=90$ improves in terms of beamwidth. The end-fire mode with the theoretical Hansen-Woodyard condition does not follow a clear trend. It is worth noting that other $\Delta\Phi$ values might provide better performance, but we restrict our exploration to cases with simple phase shifters.

\begin{table}[!t] 
\caption{Performance of scaled arrays within a realistic chip package} 
\vspace{-0.2cm}
\label{tab:scalability}
\centering
\footnotesize
\begin{tabular}{|c|c|c|c|c|} 
\hline
Size & Phases & Radiation Type & Max Gain & Beamwidth \\
\hline \hline
1 & 0 & Omnidirectional & -26.2 dB & 360\textsuperscript{o} \\  
2 & 0 0 & Omnidirectional & -22.9 dB & 360\textsuperscript{o} \\   
3 & 0 0 0 & Omnidirectional & -21.3 dB & 360\textsuperscript{o} \\ 
4 & 0 0 0 0 & Omnidirectional & -20.2 dB & 360\textsuperscript{o} \\ 
\hline 
2 & 0 90 & End-fire one-sided & -23.7 dB & 138.7\textsuperscript{o} \\ 
3 & 0 90 180 & End-fire one-sided & -23.5 dB & 122.4\textsuperscript{o} \\ 
4 & 0 90 180 270 & End-fire one-sided & -23.3 dB & 109.9\textsuperscript{o} \\ 
\hline 
2 & 0 180 & End-fire two-sided & -32.9 dB & 112.8\textsuperscript{o} \\ 
3 & 0 150 300 & End-fire one-sided & -30 dB & 120.3\textsuperscript{o} \\
4 & 0 135 270 45 & End-fire one-sided & -32.6 dB & 193.1\textsuperscript{o} \\
\hline 
\end{tabular}
\vspace{-0.2cm}
\end{table}

\noindent \textbf{Overhead:} for a first overhead estimation, we note that phase shifters at 60 GHz as small as 0.034 mm\textsuperscript{2} are available in 65-nm CMOS \cite{Meng2013}. The memory required at the controller is negligible compared to the large caches present in current multiprocessors. As justified in \ref{sec:arch}, we assume a 2-cycle delay and no impact on the network throughput. We leave a more thorough analysis for future work.

\begin{figure}[!t]
\centering
\vspace{-0.2cm}
\includegraphics[width=0.7\columnwidth]{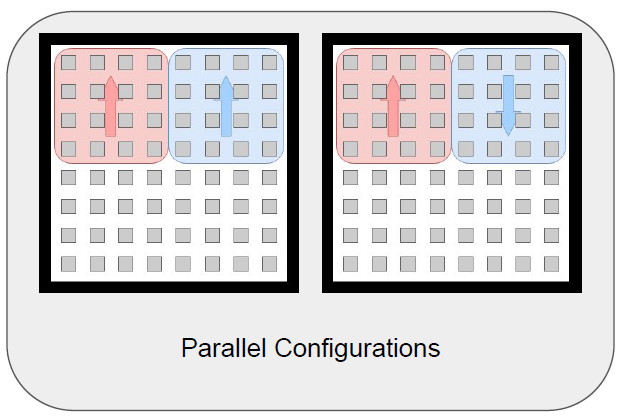}
\vspace{-0.3cm}
\caption{Example of two parallel channels created with the proposed approach.}
\vspace{-0.3cm}
\label{interf}
\end{figure}

\noindent \textbf{Spatial multiplexing:} the creation of directional beams allows to create multiple concurrent row/column channels that do not interfere each other. The study herein can be applied to develop a signal-to-interference model within the chip, through which a set of simple clustering and spatial multiplexing rules can be derived. Figure \ref{interf} shows a simple example where two independent channels can be created with directional radiation in two different columns. In future work, we plan to systematically analyze the possibilities in this respect.

\section{Conclusions}
\label{sec:conclusions}
We have presented an opportunistic scheme that leverages existing antennas in WNoC environments to create reconfigurable arrays. Albeit limited in number of beams, the proposed scheme is architecturally relevant as it can be used to implement row/column communication patterns. We simulated the feasible array configurations within a realistic chip package and found that their patterns are in close agreement with theory --although with a significantly lower efficiency due to the effects of lossy silicon and nearby antennas.

%
%
%

%
\section*{Acknowledgment}
This work was supported by ICREA under the ICREA Academia programme, the Spanish MINECO (PCIN-2015-012), the EU's H2020 FET-OPEN program (grant 736876), and the NSF (CCF 16-29431).







\begin{thebibliography}{10}
\providecommand{\url}[1]{#1}
\csname url@samestyle\endcsname
\providecommand{\newblock}{\relax}
\providecommand{\bibinfo}[2]{#2}
\providecommand{\BIBentrySTDinterwordspacing}{\spaceskip=0pt\relax}
\providecommand{\BIBentryALTinterwordstretchfactor}{4}
\providecommand{\BIBentryALTinterwordspacing}{\spaceskip=\fontdimen2\font plus
\BIBentryALTinterwordstretchfactor\fontdimen3\font minus
  \fontdimen4\font\relax}
\providecommand{\BIBforeignlanguage}[2]{{%
\expandafter\ifx\csname l@#1\endcsname\relax
\typeout{** WARNING: IEEEtran.bst: No hyphenation pattern has been}%
\typeout{** loaded for the language `#1'. Using the pattern for}%
\typeout{** the default language instead.}%
\else
\language=\csname l@#1\endcsname
\fi
#2}}
\providecommand{\BIBdecl}{\relax}
\BIBdecl

\bibitem{Jerger2017}
\BIBentryALTinterwordspacing
N.~{Enright Jerger}, T.~Krishna, and L.-S. Peh, \emph{{On-Chip Networks}},
  2nd~ed., 2017. [Online]. Available:
  \url{http://dx.doi.org/10.2200/S00209ED1V01Y200907CAC008}
\BIBentrySTDinterwordspacing

\bibitem{Bertozzi2014}
D.~Bertozzi, G.~Dimitrakopoulos, J.~Flich, and S.~Sonntag, ``{The fast evolving
  landscape of on-chip communication},'' \emph{Design Automation for Embedded
  Systems}, vol.~19, no.~1, pp. 59--76, 2015.

\bibitem{Markish2015}
O.~Markish, B.~Sheinman, O.~Katz, D.~Corcos, and D.~Elad, ``{On-chip mmWave
  Antennas and Transceivers},'' in \emph{Proceedings of the NoCS '15}, 2015, p.
  Art. 11.

\bibitem{Cheema2013}
H.~M. Cheema and A.~Shamim, ``{The last barrier: On-chip antennas},''
  \emph{IEEE Microwave Magazine}, vol.~14, no.~1, pp. 79--91, 2013.

\bibitem{Laha2015}
S.~Laha, S.~Kaya, D.~W. Matolak, W.~Rayess, D.~DiTomaso, and A.~Kodi, ``{A New
  Frontier in Ultralow Power Wireless Links: Network-on-Chip and Chip-to-Chip
  Interconnects},'' \emph{IEEE Transactions on Computer-Aided Design of
  Integrated Circuits and Systems}, vol.~34, no.~2, pp. 186--198, 2015.

\bibitem{Ahmed2018}
T.~Shinde, S.~Subramaniam, P.~Deshmukh, M.~M. Ahmed, M.~Indovina, and
  A.~Ganguly, ``{A 0.24 pJ/bit, 16 Gbps OOK Transmitter Circuit in 45-nm CMOS
  for Inter and Intra-Chip Wireless Interconnects},'' in \emph{Proceedings of
  the GLSVLSI '18}, 2018, pp. 69--74.

\bibitem{Matolak2012}
D.~Matolak, A.~Kodi, S.~Kaya, D.~DiTomaso, S.~Laha, and W.~Rayess, ``{Wireless
  networks-on-chips: architecture, wireless channel, and devices},'' \emph{IEEE
  Wireless Communications}, vol.~19, no.~5, 2012.

\bibitem{DiTomaso2015}
D.~DiTomaso, A.~Kodi, D.~Matolak, S.~Kaya, S.~Laha, and W.~Rayess, ``{A-WiNoC:
  Adaptive Wireless Network-on-Chip Architecture for Chip Multiprocessors},''
  \emph{IEEE Transactions on Parallel and Distributed Systems}, vol.~26,
  no.~12, pp. 3289--3302, 2015.

\bibitem{Mansoor2015a}
N.~Mansoor, P.~J.~S. Iruthayaraj, and A.~Ganguly, ``{Design Methodology for a
  Robust and Energy-Efficient Millimeter-Wave Wireless Network-on-Chip},''
  \emph{IEEE Transactions on Multi-Scale Computing Systems}, vol.~1, no.~1, pp.
  33--45, 2015.

\bibitem{Rezaei2016}
A.~Rezaei, M.~Daneshtalab, M.~Palesi, and D.~Zhao, ``{Efficient
  Congestion-Aware Scheme for Wireless on-Chip Networks},'' \emph{Proceedings
  of the PDP '16}, pp. 742--749, 2016.

\bibitem{Deb2013}
S.~Deb, K.~Chang, X.~Yu, S.~P. Sah, M.~Cosic, P.~P. Pande, B.~Belzer, and
  D.~Heo, ``{Design of an Energy Efficient CMOS Compatible NoC Architecture
  with Millimeter-Wave Wireless Interconnects},'' \emph{IEEE Transactions on
  Computers}, vol.~62, no.~12, pp. 2382--2396, 2013.

\bibitem{Abadal2018}
S.~Abadal, J.~Torrellas, E.~Alarc{\'{o}}n, and A.~Cabellos-Aparicio,
  ``{OrthoNoC: A Broadcast-Oriented Dual-Plane Wireless Network-on-Chip
  Architecture},'' \emph{IEEE Transactions on Parallel and Distributed
  Systems}, vol.~29, no.~3, pp. 628--641, 2018.

\bibitem{Zhang2007}
Y.~P. Zhang, Z.~M. Chen, and M.~Sun, ``{Propagation Mechanisms of Radio Waves
  Over Intra-Chip Channels With Integrated Antennas: Frequency-Domain
  Measurements and Time-Domain Analysis},'' \emph{IEEE Transactions on Antennas
  and Propagation}, vol.~55, no.~10, pp. 2900--2906, 2007.

\bibitem{Narde2018}
R.~S. Narde, N.~Mansoor, A.~Ganguly, and J.~Venkataraman, ``{On-Chip Antennas
  for Inter-Chip Wireless Interconnections: Challenges and Opportunities},'' in
  \emph{Proceedings of the EuCAP '18}, 2018.

\bibitem{Wu2017b}
J.~Wu, A.~Kodi, S.~Kaya, A.~Louri, and H.~Xin, ``{Monopoles Loaded with
  3-D-Printed Dielectrics for Future Wireless Intra-Chip Communications},''
  \emph{IEEE Transactions on Antennas and Propagation}, vol.~65, no.~12, pp.
  6838--6846, 2017.

\bibitem{Rayess2017}
W.~Rayess, D.~W. Matolak, S.~Kaya, and A.~K. Kodi, ``{Antennas and Channel
  Characteristics for Wireless Networks on Chips},'' \emph{Wireless Personal
  Communications}, vol.~95, no.~4, pp. 5039--5056, 2017.

\bibitem{Timoneda2018b}
X.~Timoneda, A.~Cabellos-Aparicio, D.~Manessis, E.~Alarc{\'{o}}n, and
  S.~Abadal, ``{Channel Characterization for Chip-scale Wireless Communications
  within Computing Packages},'' in \emph{Proceedings of the NOCS '18}, 2018.

\bibitem{Abadal2018a}
S.~Abadal, A.~Mestres, J.~Torrellas, E.~Alarc{\'{o}}n, and
  A.~Cabellos-Aparicio, ``{Medium Access Control in Wireless Network-on-Chip: A
  Context Analysis},'' \emph{IEEE Communications Magazine}, vol.~56, no.~6, pp.
  172--178, 2018.

\bibitem{DiTomaso2011}
D.~DiTomaso, A.~Kodi, S.~Kaya, and D.~Matolak, ``{iWISE: Inter-router Wireless
  Scalable Express Channels for Network-on-Chips (NoCs) Architecture},'' in
  \emph{Proceedings of the HOTI-19}.\hskip 1em plus 0.5em minus 0.4em\relax
  IEEE, 2011, pp. 11--18.

\bibitem{Vijayakumaran2014}
V.~Vijayakumaran, M.~P. Yuvaraj, N.~Mansoor, N.~Nerurkar, A.~Ganguly, and
  A.~Kwasinski, ``{CDMA Enabled Wireless Network-on-Chip},'' \emph{ACM Journal
  on Emerging Technologies in Computing Systems}, vol.~10, no.~4, p. Art. 28,
  2014.

\bibitem{Zhao2008}
D.~Zhao and Y.~Wang, ``{SD-MAC: Design and Synthesis of a Hardware-Efficient
  Collision-Free QoS-Aware MAC Protocol for Wireless Network-on-Chip},''
  \emph{IEEE Transactions on Computers}, vol.~57, no.~9, pp. 1230--1245, 2008.

\bibitem{Mondal2016}
H.~Mondal, S.~Gade, M.~Shamim, S.~Deb, and A.~Ganguly, ``{Interference-Aware
  Wireless Network-on-Chip Architecture using Directional Antennas},''
  \emph{IEEE Transactions on Multi-Scale Computing Systems}, vol.~3, no.~3, pp.
  193--205, 2017.

\bibitem{Mineo2016}
A.~Mineo, M.~Palesi, G.~Ascia, and V.~Catania, ``{Exploiting antenna
  directivity in wireless NoC architectures},'' \emph{Microprocessors and
  Microsystems}, vol.~43, no.~6, pp. 59--66, 2016.

\bibitem{Pano2017}
V.~Pano, Y.~Liu, I.~Yilmaz, A.~More, B.~Taskin, and K.~Dandekar, ``{Wireless
  NoCs Using Directional and Substrate Propagation Antennas},'' in
  \emph{Proceedings of the ISVLSI '17}, 2017, pp. 188--193.

\bibitem{Gade2018}
S.~H. Gade, S.~S. Rout, and S.~Deb, ``{On-Chip Wireless Channel Propagation :
  Impact of Antenna Directionality and Placement on Channel Performance},''
  \emph{Proceedings of the NOCS '18}, 2018.

\bibitem{Liu2016}
Z.~Liu, Y.~Liang, N.~Li, G.~Feng, H.~Yu, and S.~Chen, ``{An Energy-efficient
  Adaptive Sub-THz Wireless Interconnect with MIMO-Beamforming between Cores
  and DRAMs},'' in \emph{Proceedings of the NANOCOM '16}, 2016, pp. 1--6.

\bibitem{Baniya2018a}
P.~Baniya, S.~Yoo, K.~L. Melde, A.~Bisognin, and C.~Luxey, ``{Switched-Beam
  60-GHz Four-Element Array for Multichip Multicore System},'' \emph{IEEE
  Transactions on Components, Packaging and Manufacturing Technology}, vol.~8,
  no.~2, pp. 251--260, 2018.

\bibitem{Timoneda2018ADAPT}
\BIBentryALTinterwordspacing
X.~Timoneda, S.~Abadal, A.~Franques, D.~Manessis, J.~Zhou, J.~Torrellas,
  E.~Alarc{\'{o}}n, and A.~Cabellos-Aparicio, ``{Engineer the Channel and Adapt
  to it: Enabling Wireless Intra-Chip Communication},'' \emph{arXiv preprint
  arXiv:1901.04291}, 2018. [Online]. Available:
  \url{https://arxiv.org/pdf/1901.04291.pdf}
\BIBentrySTDinterwordspacing

\bibitem{AbadalASPLOS}
S.~Abadal, E.~Alarc{\'{o}}n, A.~Cabellos-Aparicio, and J.~Torrellas, ``{WiSync:
  An Architecture for Fast Synchronization through On-Chip Wireless
  Communication},'' in \emph{Proceedings of the ASPLOS '16}, 2016, pp. 3--17.

\bibitem{Fernando2019}
V.~Fernando, A.~Franques, S.~Abadal, S.~Misailovic, and J.~Torrellas,
  ``{Replica: A Wireless Manycore for Communication-Intensive and Approximate
  Data},'' in \emph{Proceedings of the ASPLOS '19}, 2019.

\bibitem{grama2003introduction}
A.~Grama, V.~Kumar, A.~Gupta, and G.~Karypis, \emph{Introduction to parallel
  computing}.\hskip 1em plus 0.5em minus 0.4em\relax Pearson Education, 2003.

\bibitem{Jerger2008}
N.~{Enright Jerger}, L.-S. Peh, and M.~Lipasti, ``{Virtual Circuit Tree
  Multicasting: A Case for On-Chip Hardware Multicast Support},'' in
  \emph{Proceedings of the ISCA-35}, 2008, pp. 229--240.

\bibitem{Krishna2011}
T.~Krishna, L.-S. Peh, B.~Beckmann, and S.~K. Reinhardt, ``{Towards the ideal
  on-chip fabric for 1-to-many and many-to-1 communication},'' in
  \emph{Proceedings of the MICRO-44}, 2011, pp. 71--82.

\bibitem{Park2012a}
S.~Park, T.~Krishna, C.-H. Chen, B.~Daya, A.~Chandrakasan, and L.-S. Peh,
  ``{Approaching the theoretical limits of a mesh NoC with a 16-node chip
  prototype in 45nm SOI},'' in \emph{Proceedings of the DAC-49}, 2012, pp.
  398--405.

\bibitem{BalanisBOOK}
C.~A. Balanis, \emph{{Antenna Theory: Analysis and Design}}, 3rd~ed., Wiley,
  Ed., 2005.

\bibitem{CST}
\BIBentryALTinterwordspacing
``{CST Microwave Studio}.'' [Online]. Available: \url{http://www.cst.com}
\BIBentrySTDinterwordspacing

\bibitem{Meng2013}
F.~Meng, K.~Ma, K.~S. Yeo, S.~Xu, C.~{Chye Byoon}, and W.~{Meng Lin},
  ``{Miniaturized 3-bit Phase Shifter for 60-GHz Phased-Array in 65-nm CMOS
  Technology},'' \emph{IEEE Microwave and Wireless Components Letters},
  vol.~24, no.~1, pp. 50--52, 2013.

\end{thebibliography}
\end{document}